# Modification of Thermal Conductivity of PMMA and PC by making their Nanocomposites with Carbon Nanotubes

Rajiv Bhandari[1], Neha Dhiman[2], Chetna Bajaj[2], Prashant Jindal[3], Keya Dharamvir[2] and V.K.Jindal[2*]

[1] *Post Graduate Govt. College, Sector 11, Chandigarh-160011, India*

[2] *Department of Physics, Panjab University, Chandigarh-160014, India*

[3] *University Institute of Engineering & Technology, Panjab University, Chandigarh-160014, India*

**ABSTRACT:-**Polymers – Poly methyl methacrylate (PMMA) and Poly carbonate (PC) are wonderful low cost materials which can be easily tailored and shaped. However they have poor mechanical, thermal and electrical properties which are required to be enhanced in several applications where along with high strength, a quick heat transfer becomes a necessity. Carbon nanotubes (CNT) are excellent new materials having extraordinary mechanical and transport properties. In this paper we report results of fabricating composites of varying concentrations of CNTs with PMMA and PC and measurements of thermal conductivity data by a simple transient heat flow. The samples in disk shapes of around 2 cm diameters and 0.2 cm thickness with CNT concentrations varying up to 10 wt% were fabricated. By keeping one end of the discs at steam temperature, the temperature of the other end was noted for each sample after 10 s. The rise in temperature was correlated with thermal conductivity which was appropriately calibrated. We found that both PMMA and PC measured high thermal conductivity with increase in the concentration of CNTs. The thermal conductivity of PMMA rose from about 0.2 W/mK to 0.4 W/mK at 10 wt% of CNT whereas for PC, it rose from about 0.2 W/mK to 0.9 W/mK at 10 wt% of CNT. It is thus observed that modification in thermal properties is easily achieved by making CNT based composites using only up to 10 wt% of CNTs in PMMA and PC and enabling quicker heat dissipation in these materials.

[*] Corresponding author: jindal@pu.ac.in

**I. INTRODUCTION**

Modern materials are tailored to have desired properties. Most of the research activity in today's world is devoted to designing new materials which are cost effective and easily workable. Polymethylmethacrylate (PMMA),and Polycarbonate (PC) have found great deal of applications as these are low cost and easy to work upon materials which can be changed easily for shaping. However these suffer from deficiencies depending upon the use. They are weak mechanically and low in thermal conductivity. Thermal conductivity of materials is of immense significance in the applications like power electronics, electric motors and generators, dentistry, heat exchangers, satellite devices, electronic packaging and encapsulations, mechanical gears, rotors etc. where heat gets generated and quick heat dissipation is important.

Addition of filler materials has been in use to composites with polymers with the aim to achieve thermal conductivity approximately from 1 to 30 W/mK. Traditionally, addition of thermally conductive fillers have been used such as graphite, carbon black, carbon fibres, ceramic or metal particles etc. [1]–[3]. It has been observed that such materials require high filler loadings (>30 vol.%) to achieve the level of thermal conductivity which becomes practically useful. Such high concentrations significantly alter the properties of the host material.

In recent years new carbon based nanomaterials called carbon nanotubes have been synthesised [4] which have been found to have unique electrical, mechanical and thermal properties. It is interesting to observe that their thermal conductivity is extremely high (in the range of ~1000W/mK) whether in single walled nanotubes(SWNTs) or multiwalled formation(MWNTs). The measurement of thermal conductivity of carbon nanotubes has been well described by Han and Fina [1]. Therefore we considered it worthwhile to use carbon nanotubes(CNTs) in varying concentrations in host materials in polymers (PMMA and PC). Although, thermal conductivity has already been investigated for such nano-composite systems [5], [6] but so far detailed study of the variation of thermal behaviour of nanocomposites with varying concentration and temperature has not been done especially using PMMA and PC as host material. We study two polymer materials - PMMA and PC as base matrix in the nanocomposites where CNTs are used as a filler of various concentrations.

**II SAMPLE FABRICATION**

Composites of PMMA/MWNTs with 1 wt%, 2 wt%, 5 wt%, 10 wt% (MWNTs in PMMA) and PC/MWCNTs 2.5wt%, 5wt%, 10 wt% (MWNTs in PC) were fabricated using a simple solution technique [7]–[9]. Chloroform was used as a solvent for dissolving both PMMA and PC beads with the MWNTs. The ultra-sonicator was used for random dispersion. These samples were prepared without external pressure and the alignment of MWNTs was considered random as no effort was made to align MWNTs. Subsequently, the composite material thus obtained was subjected to heat and pressure treatment in a dye (shown in Fig.1) so as to get disc shaped samples of discs with diameter 2 cm and thickness 0.2 cm.

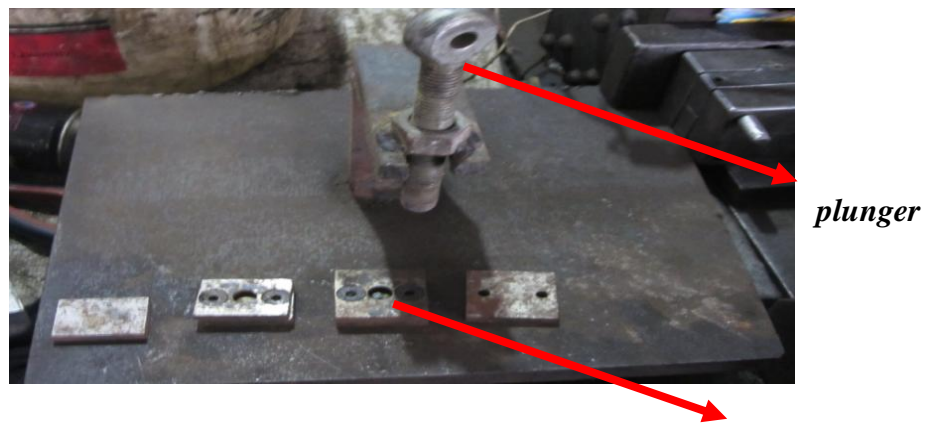

*Fig.1 setup for pallet formation of the composites* **pallet making dye**

**III THERMAL CONDUCTIVITY MEASUREMENTS**

Thermal conductivity (κ) is defined by rate of heat flow $\frac{dQ}{dt}$ through a sample of cross sectional area A having thickness $\Delta x$ with temperature gradient $\frac{\Delta T}{\Delta x}$ as:

$$\frac{dQ}{dt} = -\kappa A \frac{\Delta T}{\Delta x} \qquad (1)$$

If there are no sources of heat, we can re-write Eq.1 for net heating of an element $\Delta x$ of the specimen thickness as

$$\frac{\partial Q}{\partial t} = -\kappa A \left[ \frac{\partial T(x)}{\partial x} - \frac{\partial T(x+\Delta x)}{\partial x} \right] \quad (2)$$

This results in

$$\frac{\partial Q}{\partial t} = \kappa A \Delta x \frac{\partial^2 T}{\partial x^2} \quad (3)$$

Further, if the incremental mass of thickness Δx is heated, its x dependent rate of temperature increase is given now in terms of the density ρ and its specific heat S by

$$\frac{\rho S}{\kappa} \frac{\partial T}{\partial t} = \frac{\partial^2 T}{\partial x^2} \quad (4)$$

Which can be solved by separation of variables method,

$$T(x,t) = \Theta(t) X(x) \quad (5)$$

Giving a solution of the form

$$T(x,t) = A e^{-(\omega^2 \kappa/\rho S)t} \cos(\omega x) \quad (6)$$

Where the constants A and ω can be fixed.

In our case, there is a source of heat which maintains the hot end at a constant temperature, $T_h$. We are interested in only the time dependence of the cold end which begins to rise from $T_R$ (room temperature) to a maximum of $T_h$. Therefore we only pick up the relevant time dependence of the temperature of cold end ($T_c$) as

$$T_c = T_R + (T_h - T_R)(1 - e^{-(\omega^2 \kappa/\rho S)t}) \quad (7)$$

Therefore, under small time measurements of equal time, the thermal conductivity of two identical samples can be compared by

$$\frac{T_c - T_R}{T_h - T_R} \cong (\omega^2 \kappa/\rho S) t \quad (8)$$

$$\frac{\kappa_1}{\kappa_2} = \frac{\Delta T_{c1}}{\Delta T_{c2}} \quad (9)$$

Where $\Delta T_{c1}$ is the rise in temperature of the cold end in a small time t.

Therefore thermal conductivity can be measured by keeping one end of the disc shaped sample at a constant high temperature and measuring rise in the temperature of other end in a given time. This rise in temperature of other end will be proportional to rate of heat flow. Under the assumption that all the samples have same cross sectional area and same thickness (they are prepared in the same dye), the temperature rise of the cold end under constant time is a measure of thermal conductivity.

The average temperature increase in all samples whose one end is steam temperature, 100°C.and the other end heats up marginally above room temperature in time of a few seconds. Eq.2 was accordingly corrected for variations of mass, area and thickness, assuming specific heat remains identical as follows:

$$\frac{\kappa_1}{\kappa_2} = \frac{\rho_2}{\rho_1}\frac{\Delta T_{c1}}{\Delta T_{c2}}$$

We fabricated the setup consisting of aluminium cylindrical tube, having inlet and outlet for the passage of constant circulation of steam as shown in Fig. 2.

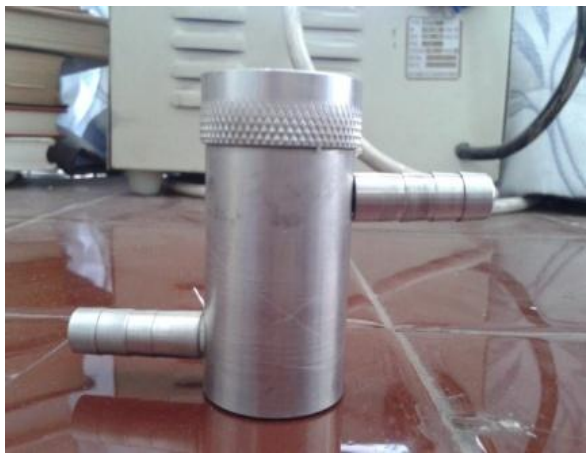

*Fig. 2 Setup for measuring Conductivity*

The each composite sample was placed on the top end of the setup and the rise in temperature on the other end was measured after 10 seconds at room temperature. Knowing thermal conductivity of pure PMMA or PC, the thermal conductivities of various samples can thus be calculated.

## IV RESULTS AND DISCUSSION

The thermal conductivities thus calculated from the observations of PMMA/MWNT with the concentration change from 1 wt% to 10 wt% of MWNT varies from 0.240 W/mK to 0.446 W/mK and has been plotted in Fig.3. For this, the value of thermal conductivity of pure PMMA, has been taken to be 0.21 W/mK [1]. There is a consistent increase in thermal conductivity by increasing the concentration of MWNTs. The average increase in thermal conductivity is around 0.02 W/mK for every 1 wt% increase in concentration of MWNT in PMMA/MWNT.

Similarly, we plot in Fig.4 the results of composites of PC/MWNTs.

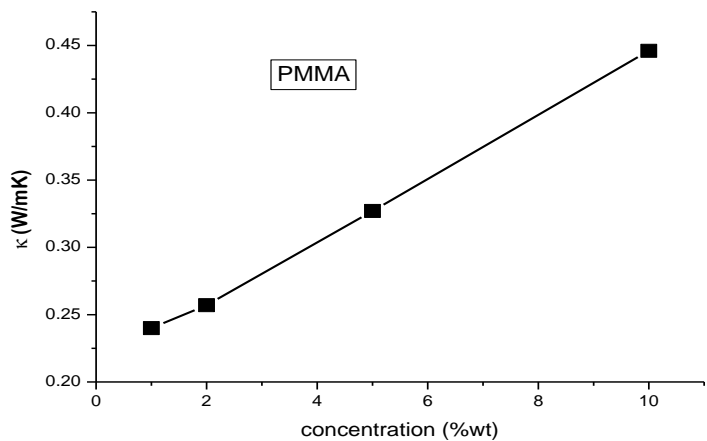

*Fig. 3 Variation of thermal conductivity with %wt concentration of MWNT in PMMA/MWNT*

Like above, Fig. 4 depicts the thermal conductivities of PC/MWCNT varies from 0.285 W/mK to 0.91 W/mk with the concentration from 2.5 wt% to 10 wt% of MWCNT in comparison to 0.19 W/mK for pure PC. It is observed that at low concentration of wt% of MWCNT in PC/MWCNT the rate of increase of thermal conductivity is 0.05 W/mK with 1 wt% rise in concentration but doubles for concentration above 5 wt%.

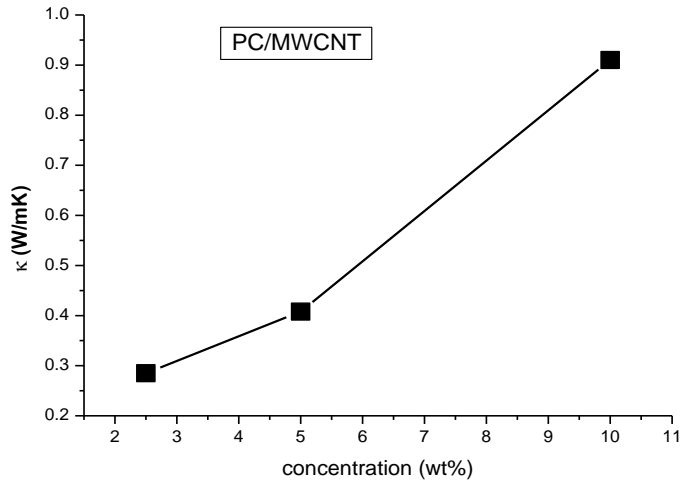

*Fig. 4 Variation of thermal conductivity with %wt concentration of MWNT in PC/MWNT*

It is found that both PMMA and PC measured high thermal conductivity with increase in the concentration of MWNTs. The thermal conductivity of PMMA/MWNT rose from about 0.2 W/mK to 0.4 W/mK at 10 wt% of MWNT whereas for PC/MWNT, it rose from about 0.2 W/mK to 0.9 W/mK at 10 wt% of MWNT. It is thus observed that modification in thermal properties is easily achieved by making MWNT based composites using only up to 10 wt% of MWNTs in PMMA and PC. This enables quicker heat dissipation in these composites.

## V  CONCLUDING REMARKS

This paper reports results on the modification of thermal conductivity of PMMA and PC composites with MWNTs. The concentration of MWNTs was varied and no efforts was made to orient nanotubes in any preferred direction. Thus the composites made were made with random orientation of MWNTs. The thermal conductivity of the composites so formed was measured and found to increase nearly factor of 2 for PMMA composites and by nearly a factor of 4 for PC composites when the concentration of MWNTs was raised to 10%. The thermal conductivities of the composites seem to represent weighted average of the thermal conductivities of MWNTs and PMMA or PC. On the basis of the result of this work, it can be concluded that MWNT play significant role in enhancing thermal conductivity of poor thermal conductors. More importantly low concentration of MWNT is adequate to increase thermal conductivity in contrast to requirement of more than 30% of non carbon nanotubes based filler materials. Thus at these low concentration we can retain the advantageous properties of the host material.